\documentclass[fleqn,12pt,twoside]{article}
\usepackage[headings]{espcrc1}
\readRCS
$Id: espcrc1.tex,v 1.2 2004/02/24 11:22:11 spepping Exp $ 
\usepackage{graphicx,epsfig,colortbl}
\usepackage[figuresright]{rotating}
\usepackage{wrapfig}

\newcommand{\AmS}{{\protect\the\textfont2
  A\kern-.1667em\lower.5ex\hbox{M}\kern-.125emS}}
\newcommand{\grb}{\multicolumn{1}{>{\columncolor[gray]{.5}}c}{}}
\newcommand{\gra}{\multicolumn{1}{>{\columncolor[gray]{.8}}c}{}}

\def\la{\langle}
\def\ra{\rangle}
\def\beq{\begin{equation}}
\def\eeq{\end{equation}}
\def\be{\begin{eqnarray}}
\def\ee{\end{eqnarray}}

\def\k2av{\la k_T^2\ra}

\title{Molecular dynamics simulation of strongly coupled QCD plasmas }

\author{
P. Hartmann\address[SZFKI]{Research Institute for Solid State Physics and Optics of the Hungarian Academy of Sciences, P.O. Box 49, Budapest H-1525, Hungary},
Z. Donk\'o\addressmark[SZFKI],
P. L\'evai\address[RMKI]{RMKI Research Institute for
        Particle and Nuclear Physics, \\
        P.O. Box 49, Budapest H-1525, Hungary},
and G. J. Kalman\address[Boston]{Department of Physics, Boston College, Chestnut Hill, Massachusetts 02467, USA},
\thanks{We thank Mikl\'os Gyulassy and Ferenc Szalai for helpful discussions.
This work was supported in part by  U.S. DOE grant DE-FG02-03ER5471, NFS grants PHYS-0206695, PHYS-0514619 and Hungarian grants OTKA T-043455, T-48389, PD-049991, MTA-OTKA 90/46140.}}

\begin{document}
\date{15 September 2005} 
\maketitle

\begin{abstract}
The properties of a strongly interacting quark plasma are
investigated by molecular dynamics method including non-abelian
quark-quark potential.
Our main goal is to study the thermalization process
in this system. We find an interesting resonance-like
behaviour: at a characteristic time close to the inverse plasma frequency
the quark plasma is heated up substantially via energy transfer
from quark potential energy into one particle kinetic energy. 
Color rotation mechanism enhances the effectivity of this heating process, 
leading to a very fast thermalization with high temperature.


\end{abstract}

\section{ Introduction}

Early theoretical speculations suggested the appearance of 
a weakly interacting perturbative quark-gluon plasma state 
in heavy ion collisions at ultrarelativistic energies. However, 
recent data obtained at RHIC experiments, and their theoretical 
analysis, indicate the formation of a strongly interacting 
deconfined matter \cite{sQCD}. 
In fact, lattice-QCD calculations on equation of state have 
already shown the strong coupling and the theoretical analysis 
based on massive quasi-particle picture described quantitatively 
many properties of this matter \cite{QCDquasi}. 
Thus we expect a relatively dilute, massive, quark and antiquark 
dominated matter to appear before hadronization.  
Widely used molecular dynamics simulation \cite{MD,OCP} 
is the ideal tool to study such a particle compound, to investigate 
non-perturbative features and to determine different properties
in the strongly interacting matter. 
In this simulation we can study the influence of the color 
charge fluctuation and  pair correlation function can be generated, 
displaying if the quark matter is gaseous or 
rather liquid like. Although molecular dynamics models for 
non-abelian interactions are very much approximate, we can 
study qualitatively very important microscopic processes,
for what other methods do not exist. 
Here we summarize the main findings of our work, details can be found 
in Ref. \cite{long}.

\section{The Molecular Dynamics Model}

In the model the quarks are the only actively simulated species,
the gluons are represented as a background field. This approach is similar to
the case of the classical multicomponent or one component plasma (OCP) model, where the
motion of the ions is of interest, the electronic background is
treated as a constant field \cite{OCP,Thoma}.

In the model pairwise interaction of the quarks is assumed. The
possible realizations of (single flavored) two-quark (QQ) system are composed of symmetric sextet and antisymmetric anti-triplet combinations.

\begin{wraptable}{r}{0.4 \textwidth}
  \vspace{-1.3cm}
  \caption{\footnotesize $\langle\lambda_i \cdot \lambda_j\rangle$ values for QQ-pairs.}
  \label{table:2}
  \renewcommand{\tabcolsep}{1pc} 
  \renewcommand{\arraystretch}{1.2} 
  \begin{tabular}{c|c|c|c}
  & $|R\rangle$  & $|G\rangle$  & $|B\rangle$ \\
  \hline
  $|R\rangle$ & $+\frac{1}{3}$  & $D$ & $D$ \\
  \hline
  $|G\rangle$ & $D$  & $+\frac{1}{3}$ & $D$ \\
  \hline
  $|B\rangle$ & $D$  & $D$ & $+\frac{1}{3}$ \\
  \hline
  \end{tabular} 
  \vspace{-1.7cm}
\end{wraptable}

The quark-quark effective interaction is represented by a Coulomb-like potential in the form of
\begin{equation}
V=\langle\lambda_i \cdot \lambda_j\rangle\frac{\alpha}{r_{ij}}
\label{eq:V}
\end{equation}
acting between the $i^{th}$ and $j^{th}$ particle, where
$\alpha = g^2/4\pi$ is the interaction coupling constant, $r_{ij}$ is
the inter-particle distance and $\langle\lambda_i \cdot \lambda_j\rangle$
depends on the color states of the particles as shown in
Table~\ref{table:2}. The parameter $D$ introduced in
Table~\ref{table:2} represents values of $\langle\lambda_i \cdot
\lambda_j\rangle = +\frac{1}{3}$ or $-\frac{2}{3}$ randomly chosen
with equal probabilities, due to the random realization of
symmetric and antisymmetric QQ-states for all different colored
pairs. 
This has the effect, that equally colored quarks interact
always repulsively, but different colors may attract or repell each
other. An additional short-distance cutoff
in the interaction is introduced: at $r<0.1$~fm the potential is taken
as a linear function of $r$, which results in a smooth quadratic decrease of the inter-particle force for $r \rightarrow 0$ 
(soft-core model). This arbitrary cutoff can be justified in view of
the finite spatial extension of the quark wave-function.

\begin{wraptable}{r}{0.5 \textwidth}
  \vspace{-1cm}
  \caption{\footnotesize Interaction table example for a 9-quark system. Dark and light gray fields are excluded due to the neglected self interaction and double counting, respectively.}
  \label{table:3}
  \renewcommand{\tabcolsep}{0.3pc} 
  \renewcommand{\arraystretch}{1.3} 
  \smallskip
  \begin{tabular}{c|c|c|c|c|c|c|c|c|c}
  & $1_R$  & $2_G$  & $3_B$ & $4_G$  & $5_R$  & $6_R$ & $7_B$  & $8_G$ & $9_B$\\
  \hline
  $1_R$ & \grb & $D$ & $D$ & $D$ & $+\frac{1}{3}$ & $+\frac{1}{3}$ & $D$ & $D$ & $D$\\
  \hline
  $2_G$ & \gra & \grb & $D$ & $+\frac{1}{3}$ & $D$ & $D$ & $D$ & $+\frac{1}{3}$ & $D$\\
  \hline
  $3_B$ & \gra & \gra & \grb & $D$ & $D$ & $D$ & $+\frac{1}{3}$ & $D$ & $+\frac{1}{3}$\\
  \hline
  $4_G$ & \gra & \gra & \gra & \grb & $D$ & $D$ & $D$ & $+\frac{1}{3}$ & $D$\\
  \hline
  $5_R$ & \gra & \gra & \gra & \gra & \grb & $+\frac{1}{3}$ & $D$ & $D$ & $D$\\
  \hline
  $6_R$ & \gra & \gra & \gra & \gra & \gra & \grb & $D$ & $D$ & $D$\\
  \hline
  $7_B$ & \gra & \gra & \gra & \gra & \gra & \gra & \grb & $D$ & $+\frac{1}{3}$\\
  \hline
  $8_G$ & \gra & \gra & \gra & \gra & \gra & \gra & \gra & \grb & $D$\\
  \hline
  $9_B$ & \gra & \gra & \gra & \gra & \gra & \gra & \gra & \gra & \grb \\
  \hline
  \end{tabular}\\[2pt]
  \vspace{-1.3cm}
\end{wraptable}

The bookkeeping of all possible QQ pair-interactions in the system is realized using the so called ``interaction table'' of which an example is shown in table~\ref{table:3}. The particles are assigned a label and all particles have an associated color value. The elements of the table show the $\langle\lambda_i \cdot \lambda_j\rangle$ values of the actual QQ pair realizations. The values of the $D$ entries are randomly re-distributed periodically (see below). The total system is color-balanced (white), but neglecting the self-interaction results in the appearance of a net attraction.

In the molecular dynamics simulation the quarks interact with the color-dependent
effective pair-potential given by (\ref{eq:V}). Periodic boundary
conditions are applied together with the Ewald summation technique for the proper
treatment of the long range interaction. Forces acting on each of
the particles (quarks) due to all other particles are calculated in
every timestep and the equation of motion for each particle
is integrated in time \cite{MD}.

At the start of the simulation the particles are randomly distributed
in the cubic simulation box with kinetic energies determined by an
initial temperature. An initialization phase is used before any measurement is taken on the system. In this phase the momenta of the particles are scaled back according to the given initial temperature in every timestep. The simulation time of this phase is chosen to be long enough to reach equilibrium conditions for the unperturbed system over the subsequent measurement period.

The effect of the gluon field on the quark component is taken into
account in three ways: (i) it represents an infinite heat reservoir,
energy can be coupled out of it without any changes in the qluon
field; (ii) gluons can transfer color charge, the color state of two
quarks may be exchanged (color rotation) with a characteristic time
given by $\tau_C$; (iii) qluons perturb QQ-pairs, leading to a new
realizations of the pair-interaction (redistribution of the
interaction table) with a characteristic time $\tau_D$.

The input parameters in our simulations are: density
$n=10$~quarks/fm$^3$, initial temperature $T_0=200$~MeV, effective
quark mass $m_q=300$~MeV/c$^2$, interaction
coupling strength $\alpha=1$, time between reassignment of all $D$
(different-color pair elements) in the interaction table
$\tau_D$, time between color rotation events
$\tau_C$ and the ratio of pairs color rotated in one event
$N_C$. The plasma frequency
is derived as $\omega_P^2=4\pi\alpha n/m_q$. The behavior of the system is now governed by the two, independently chosen ``refreshment'' times $\tau_D$ and $\tau_C$.

\begin{wrapfigure}{l}{0.6 \textwidth}
\vspace{-0.6cm}
\includegraphics[scale=0.75]{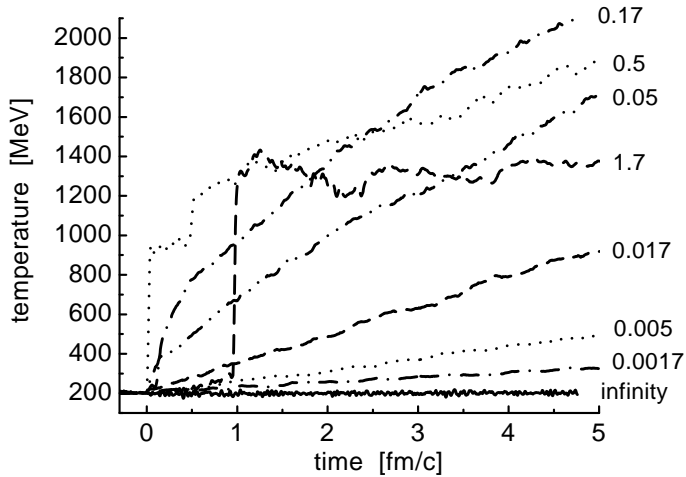}
\caption{ \footnotesize Time-evolution of the kinetic temperature for systems with different $\tau_D$ values at $\tau_C=0$. (Numbers labeling the curves are the corresponding $\tau_D$ values in fm/c units.)}
\label{fig1}
\vspace{-0.6cm}
\end{wrapfigure}     

 Figure~\ref{fig1} shows the time-evolution of the quark kinetic temperature  for different $\tau_D$ values. It can be seen, that in the limiting cases $\tau_D = \infty$ and $\tau_D \rightarrow 0$ the temperature fluctuates closely to its initial value. Intermediate values of $\tau_D$ result in increasing temperatures. 

\begin{figure}[htb]
\begin{center}
\includegraphics[scale=0.75]{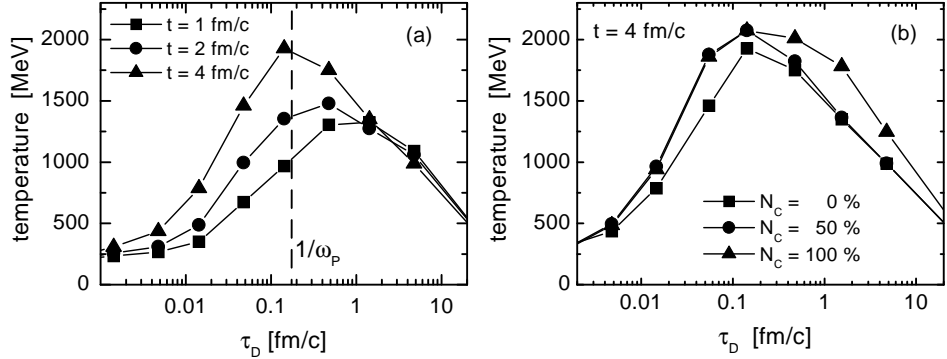}
\vspace{-0.8cm}
\caption{ \footnotesize Temperature vs. $\tau_D$ at (a) times $t=1$, 2 and 4~fm/c without color rotation; and (b) at time $t=4$~fm/c with $\tau_C=\tau_D$ for color-rotation ratios $N_C=0$, 50 and 100\%.}
\label{fig2}
\end{center}
\vspace{-1cm}
\end{figure}

The $\tau_D$-dependence of the temperature is plotted in figure~\ref{fig2}(a) for three snapshots in time. A remarkable resonant-like behavior of the heating of the system can be observed, where the maximum heating-rate establishes at $\tau_D$ values around the inverse plasma frequency $1/\omega_P \approx 0.19$~fm/c.
Figure~\ref{fig2}(b) shows the dependence of the resonant-like peak on an additional color rotation rate, where $\tau_C=\tau_D$. We have found that the color exchange increases the energy transfer from the background field to the quark component by 10 to 20\%. 

\begin{figure}[htb]
\begin{center}
\includegraphics[scale=0.75]{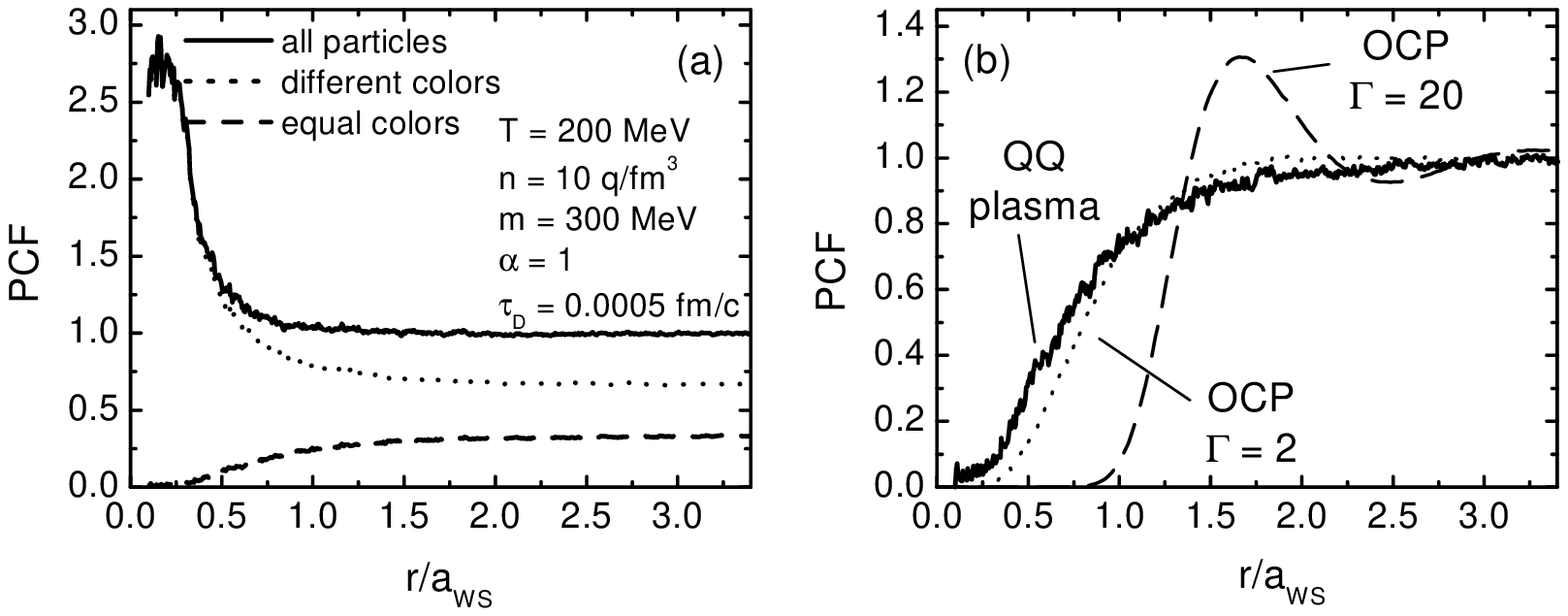}
\vspace{-0.8cm}
\caption{ \footnotesize Pair-correlation function for the $\tau_D = 0.0005$~fm/c case (a) decomposed into equal-colored and different-colored contributions; and (b) compared to classical OCP plasmas. Here $a_{WS}=0.3628$~fm is the Wigner-Seitz radius, and $\Gamma$ is the plasma (non-ideality) parameter.}
\label{fig3}
\end{center}
\vspace{-1cm}
\end{figure}

Figure~\ref{fig3}(a) displays the pair-correlation function (PCF) for the $\tau_D=0.0005$~fm/c case neglecting color-rotation. The total PCF is decomposed into contributions of equal-colored-, and different-colored pairs. The attractive force acting between some of the different-colored pairs result in the appearance of the correlation peak at $r \approx 0$. This indicates the onset of clusterization \cite{Shuryak,Fisher} of particles which gets more pronounced at higher $\tau_D$ values. Focusing only on the (always) repulsive equal-colored pairs, the PCF can be compared with classical OCP results [see figure~\ref{fig3}(b)]. This comparison shows that the repulsive equal-colored component of the QQ plasma is structurally similar to classical OCP in the so called ``gas'' phase (at $\Gamma$ values around unity), which indicates already strong non-ideality of the QQ-plasma investigated here.

\begin{thebibliography}{99}

\bibitem{sQCD} 
M. Gyulassy and L.McLerran, Nucl. Phys. A {\bf 750}, 30 (2005).

\bibitem{QCDquasi} 
P. L\'evai, U. Heinz, Phys. Rev. C {\bf 57}, 1879 (1998).

\bibitem{MD} D. Frenkel and B. Smit, {\it Understanding Molecular
    Dynamics Simulations} (Academic Press, New York, 2001).

\bibitem{OCP} P. Hartmann, G. J. Kalman, Z. Donk\'o and K. Kutasi,
  Phys. Rev. E {\bf 72}, 026409 (2005).

\bibitem{long}
P. Hartmann, Z. Donk\'o, G. J. Kalman and P. L\'evai, to be published.

\bibitem{Thoma}
M. Thoma, these proceedings.

\bibitem{Shuryak}
E. V. Shuryak and I. Zahed, Phys. Rev. C {\bf 70}, 021901(R) (2004). 

\bibitem{Fisher}
M. E. Fisher and Y. Levin, Phys. Rev. Lett. {\bf 71}, 3826 (1993).



\end{thebibliography}
\end{document}